\documentstyle[12pt]{article}

\normalsize

\def\a{\alpha}
\def\b{\beta}

\def\d{\delta}

\def\g{\gamma}

\def\l{\lambda}
\def\m{\mu}

\def\p{\pi}

\def\D{\Delta}

\def\G{\Gamma}

\def\cd{{\cal D}}

\def\cp{{\cal P}}

\def\cu{{\cal U}}

\def\iff{\Leftrightarrow}

\def\bar#1{\overline{#1}}

\def\Hat#1{\rlap{\kern.10em$\widehat{\phantom G}$}#1}
\def\HAt#1{\rlap{\kern.05em$\widehat{\phantom G}$}#1}

\def\cAp#1{\rlap{\kern.1em$\widehat{\phantom{G\vrule height.8em}}$}#1{}}
\def\Cap#1{\rlap{\kern.05em$\widehat{\phantom{G\vrule height.8em}}$}#1{}}

\let\oldtheequation=\theequation
\def\doteqs#1{\setcounter{equation}{0}
	    \def\theequation{{#1}.\oldtheequation}}
\newcounter{sxn}
\def\sx#1{\addtocounter{sxn}{1} \bigskip\medskip \goodbreak
\noindent{\large\bf\centerline{\thesxn.~~#1}} \nobreak \medskip}
\def\sxn#1{\sx{#1} \doteqs{\thesxn}}

\newcounter{axn}

\def\br{\begin{thebibliography}{abc}}
\def\er{\end{thebibliography}}
\def\rf{\bibitem}
\date{}

\begin{document}

\bibliographystyle{unsrt}
\footskip 1.0cm
\thispagestyle{empty}
\setcounter{page}{0}
\begin{flushright}
Napoli: DSF-T-29/93, INFN-NA-IV-29/93\\
Syracuse: SU-4240-550\\
October 1993\\
\end{flushright}
\vspace{10mm}

\centerline {\LARGE FINITE APPROXIMATIONS TO QUANTUM PHYSICS:}
\vspace{5mm}
\centerline {\LARGE QUANTUM POINTS AND THEIR BUNDLES }
\vspace*{15mm}
\centerline {\large A.P. Balachandran, G. Bimonte, E. Ercolessi }
\vspace{5mm}
\centerline{ \small and }
\vspace*{5mm}
\centerline{ \large P. Teotonio-Sobrinho}
\vspace*{5mm}
\centerline {\it Department of Physics, Syracuse University,}
\centerline {\it Syracuse, NY 13244-1130}
\vspace*{25mm}
\normalsize
\centerline {\bf Abstract}
\vspace*{5mm}

There exists a physically well motivated method for approximating manifolds
by certain topological spaces with a finite or a countable set of points.
These spaces, which are partially ordered sets (posets) have the power to
effectively reproduce important topological features of continuum physics like
winding numbers and fractional statistics, and that too often with just a
few points. In this work, we develop the essential tools for doing quantum
physics on posets. The poset approach to covering space quantization, soliton
physics, gauge theories and the Dirac equation are discussed with emphasis
on physically important topological aspects. These ideas are illustrated by
simple examples like the covering space quantization of a particle on a
circle, and the sine-Gordon solitons.

\newpage

\baselineskip=24pt
\setcounter{page}{1}
\newcommand{\be}{\begin{equation}}
\newcommand{\ee}{\end{equation}}

\sxn{Introduction}\label{sec-introduction}

All known successful fundamental models for the description of nature are
based on continuum physics. In the latter, we typically assume that spacetime
is a manifold and describe observables by classically smooth structures
painted on this background. Such models have given us spectacularly faithful
accounts of nature after quantisation, agreement between theory and experiment
for electron $g-2$ being about one part in $10 ^8$ \cite{g2}. There is at
present no empirical evidence whatsoever that spacetime at a fundamental level
is anything but a continuum.

Very good reasons nevertheless exist for studying discrete physics. Natural
laws are often exceedingly complex and their quantitative investigation
involves their discretisation and adaptation to machines. Also discrete models
have many times provided us with soluble systems which illustrate
physical principles, or approximate the more accurate continuum theory in a
satisfactory manner, well known examples being tight binding \cite{tight},
Ising \cite{ising} and Hubbard models \cite{hubbard} in chemistry and
condensed matter theory.

As another reason for the study of discrete physics, mention may also be made
of suggestions that spacetime has a discrete structure at the
Planck scale \cite{plank}.

In this paper, we address a serious limitation of conventional discretisations
of conti\-nuum theories caused by their naive treatment of topology. In
lattice QCD \cite{lattice} for example, spacetime topology is trivial with
each point forming an open set and bears no resemblance to the continuum
topology. There are similar defects, particularly harmful for topological
studies, in discretised investigations of vortices in Higgs models
\cite{vortex} or of solitons in Skyrme physics \cite{bal}. No nontrivial
winding numbers for instance exist in the usual finite approximations to
the  Skyrme soliton while it is virtually impossible to display the
qualitative bundle theoretical features of monopoles \cite{bal} in trivial
discretisations.

Some time ago, Sorkin \cite{sorkin} revived and extended an old method
\cite{aleksandrov} for the approximation of continuous spaces by finite
or countable partially ordered sets (posets). This approximation, as  we
shall see, is physically well motivated. Further the topology of posets,
derived from the continuum and approximating the latter, is not simple
minded. In particular, the homotopy and homology groups $\p_n$ and $H_n$
of posets \cite{sorkin,aleksandrov,hilton,stanley} are genericaly nontrivial
if that is
the case for their continuum limits. These approximations are often so good
that the continuum $\p_n$ and $H_n$ are reproduced with just a few points
\cite{???}.

In this paper, we initiate the development of tools for doing quantum and
soliton physics on posets. An effort has been made to keep the discussion
elementary even at the expense of avoiding important proofs or physically
significant examples. In Section 2, earlier research and especially Sorkin's
work are summarised and illustrated using low dimensional manifolds. Their
approximations by posets, and the representation of the latter by Hasse
diagrams
\cite{sorkin,stanley} are also explained. Then to show how accurate the poset
approximation can be, we examine $\p_1$ for a four point poset approximation
to the circle $S^1$ and argue that it is ${\bf Z}$, the same as for $S^1$.

Section 3 introduces a few of the basic tools, and in particular the poset
Laplacian $\D$, essential for quantum physics. The construction of the
latter involves the
definition first of a certain boundary operator $\partial $ on posets
\cite{hilton} and of its dual $d$, much as in homology and cohomology theories.
Intruducing a scalar product on functions of posets, the adjoint
$d^{\dag}$ or $\d$ of $d$ can then be defined, $\D$ being just $-d\d -\d d$
exactly as for continuum \cite{munkres}. Energy levels of a `free' particle
on circle posets are also briefly examined as an illustration.

It is well known that the energy eigenstates of a free particle on $S^1$ can
change by a phase when one goes around $S^1$, this phase being occasionally
explained with the help of a magnetic flux threading $S^1$ \cite{geons}. For
manifolds, a general approach to such possibilities involves regarding wave
functions as functions on the universal covering space $\bar Q$ of the
configuration space $Q=\bar Q/\p_1(Q)$ [$\bar Q$ being ${\bf R}^1$ for
$Q=S^1$] which transform by a unitary irreducible representation (UIR) of
$\p_1(Q)$ [$\p_1(S^1)$ being {\bf Z}] \cite{bal}. We show in Section 4 that
similar ideas work well for posets too, every poset $P$ having another simply
connected poset $\bar P$ as its universal cover
with $P=\bar P/\p_1(P)$. There is
also a Laplacian $\bar \D$ on $\bar P$. For circle posets, the spectrum of
$\bar \D$ for wave functions transforming by a UIR of $\p_1(P)$ is worked out
and shown to reproduce the flux effect previously alluded to.

Section 5 deals with topological solitons. The space $F$ of soliton fields
consists of
smooth maps of a domain space $D$ to a target space $T$ in continuum physics.
We discretise $D$ and $T$ to posets $P(D)$ and $P(T)$ and consider instead
the space of
continuous maps of $P(D)$ to $P(T)$ as approximations of $F$. We argue that
this space $P^{(0)}(F)$ has
topological sectors and can describe topological solitons, and can also be
converted to a poset. So quantum physics can be done using $P^{(0)}(F)$. The
nature of $P^{(0)}(F)$ is illustrated for $D=T=S^1$, the appropriate choice
for the sine-Gordon model \cite{bal}.

This introductory article does not discuss several inportant topics like Dirac
equation and gauge theories on posets, and the computerisation of poset
calculations, until Section 6. The latter concludes the paper by outlining
our progress in their study.

\sxn{The Finite Approximation}

Let $Q$ be a continuous topological space like for example the sphere $S^N$ or
the Euclidean space ${\bf R}^N$. Experiments are never so accurate that
they can detect events associated
with points of Q, rather they will only detect events as occurring
in certain sets $O_\l$. It is therefore
natural to identify any two points $x$, $y$ of $Q$ if every set $O_\l$
containing either point contains the other too. Let us assume that the sets
$O_\l$ cover $Q$,
\be
   Q=\bigcup _\l O_\l \label{2.1}
\ee
and write $x\sim y$ if $x$ and $y$ are not separated or distinguished by
$O_\l$ in the sense above:
\be
   x\sim y \mbox{  means  } x\in O_\l \iff y\in O_\l\,. \label{2.2}
\ee
Then $\sim $ is an equivalence relation, and it is reasonable to replace $Q$
by $Q/ \sim \equiv P(Q)$ to reflect the coarseness of observations. It is this
space, obtained by identifying equivalent points, that will be our
approximation for $Q$.

We assume that the number of $O_\l$ is finite when $Q$ is compact so that
$P(Q)$ is an approximation to $Q$ by a finite set in this case. When $Q$ is
not compact, we assume instead that each point has a neighbourhood
intersected by only finitely many $O_\l$ so that $P(Q)$ is a ``finitary"
approximation to $Q$ \cite{sorkin}. We also assume that each $O_\l$ is open
\cite{top} and that
\be
   \cu =\{O_\l \} \label{2.3}
\ee
is a topology for $Q$. This implies that $O_\l\cup O_\m$ and
$O_\l\cap O_\m$ $\in \cu$ if $O_{\l,\m}\in \cu$. Now experiments can
isolate events in $O_\l\cup O_\m$ and $O_\l\cap O_\m$ if they can do so
in $O_\l$ and $O_\m$ separately, the former by detecting an event in either
$O_\l$ or $O_\m$, and the latter by detecting it in both $O_\l$ and $O_\m$.
The hypothesis that $\cu $ is a topology is thus conceptually consistent.

These assumptions allow us to isolate events in certain sets of the form
$O_\l\setminus [O_\l\cap O_\m]$ which may not be open. This means that there
are in general points in $P(Q)$ coming from sets which are not open in $Q$.

In the notation we employ, if $P(Q)$ has $N$ points, we sometimes denote it
by $P_N(Q)$.

Let us illustrate these considerations for a cover of $Q=S^1$ by four open
sets as in Fig. 1(a).
In that figure, $O_{1,3}\subset O_2\cap O_4$.
Figure 1(b) shows the corresponding discrete space $P_4(S^1)$, the
points $x_i$ being the images of sets in $S^1$. The map $S^1\rightarrow
P_4(S^1)$ is given by
$$
  O_1 \rightarrow x_1, \;\;\;\;\;
	O_2\setminus [O_2\cap O_4] \rightarrow x_2,
$$
\be
  O_3 \rightarrow x_3, \;\;\;\;\;
	O_4\setminus [O_2\cap O_4] \rightarrow x_4\,.\label{2.3a}
\ee

Now $P(Q)$ inherits the quotient topology from $Q$\cite{top}.
It is defined as follows.
Let $\Phi $ be the map from $Q$ to $P(Q)$ obtained by identifying equivalent
points. An example of $\Phi $ is given by (\ref{2.3a}). In the quotient
topology, a set in $P(Q)$ is declared to be open if its inverse image for
$\Phi $ is open in $Q$. It is the finest topology compatible with the
continuity of $\Phi $. We adopt it hereafter as the topology for $P(Q)$.

This topology for $P_4(S^1)$ can be read off from Fig.1, the open sets being
\be
\{x_1\},\,\,\{x_3\},\,\,\{x_1,x_2,x_3\},\,\,\{x_1,x_4,x_3\},\,\,\label{2.4}
\ee
and their unions and intersections (an arbitrary number of the latter being
allowed, $P_4(S^1)$ being finite).

A partial order $\preceq $ \cite{aleksandrov,hilton,stanley} can be introduced
in $P(Q)$ by declaring that $x\preceq y$ if every open set containing $y$
contains also $x$. For $P_4(S^1)$, this order reads
\be
x_1\preceq x_2,\,\,\,x_1\preceq x_4;\,\,\,x_3\preceq x_2,\,\,\,
x_3\preceq x_4,\,\,\, \label{2.5}
\ee
where we have omitted writing the relations $x_j\preceq x_j$.

Later, we will write $x\prec y$ to indicate that $x\preceq y$ and $x\neq y$.

It is to be observerd that if $x\preceq y$, then the sequence $x,x,x,...$
converges to $y$ by the usual definition of convergence. So we may also
write $x\rightarrow y$ if $x\preceq y$.

In a Hausdorff space \cite{top}, there are open sets $O_x$ and $O_y$ for any
two distinct points $x$ and $y$ such that $O_x\cap O_y=\emptyset $. A
finite Hausdorff space has discrete topology with each point being an open
set. So $P(Q)$ is not Hausdorff. But
it is what is called $T_0$, where for any two distinct points, there is an
open set containing at least one of these points and not the other. For
$x_1$ and $x_2$ of $P_4(S^1)$, the open set $\{x_1 \}$ contains $x_1$ and
not $x_2$, but there is no open set containing $x_2$ and not $x_1$.

Any poset can be represented by a Hasse diagram constructed using the
following rules: 1) If $x\prec y$, then $y$ is higher than $x$. 2) If
$x\prec y$ and there is no $z$ such that $x\prec z\prec y$, then $x$ and $y$
are connected by a line called a link.

In case 2), $y$ is said to cover $x$.

The Hasse diagram for $P_4(S^1)$ is shown in Fig. 2.

The smallest open set $O_x$ containing $x$ consists of all $y$ converging to
$x$ ($y\rightarrow x$)[so that the closure of the singleton set $\{y\}$
contains $x$].
In the Hasse diagram, it consists of $x$ and all
points we encounter as we travel along links from $x$ to the bottom. In
Fig. 2, this rule gives
$\{x_1,x_2,x_3\}$ as the smallest open set containing $x_2$, just as in
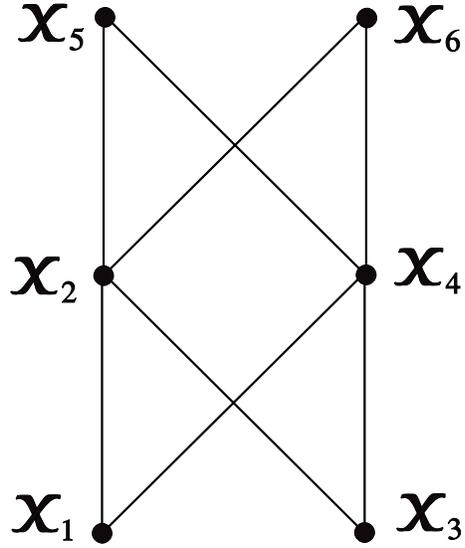
(\ref{2.4}). As another example, consider the Hasse diagram of Fig. 3
for a two-sphere
poset $P_6(S^2)$ derived in \cite{sorkin}. Its open sets are generated by
$$
\{x_1\},\,\,\,\{x_3\},\,\,\,\{x_1,x_2,x_3\},\,\,\,\{x_1,x_4,x_3\}
$$
\be
\{x_1,x_2,x_5,x_4,x_3\},\,\,\,\{x_1,x_2,x_6,x_4,x_3\},\,\,\,\label{2.6}
\ee
by taking unions and intersections.

As one more example, Fig. 4 shows a cover of $S^1$ by $2N$ open sets $O_j$ and
the Hasse diagram of its poset $P_{2N}(S^1)$.

We conclude this Section by arguing that the fundamental group $\p_1$ of
$P_4(S^1)$ is ${\bf Z}$ \cite{sorkin}.
This group is obtained from
continuous maps of $S^1$ to $P_4(S^1)$, or equivalently, of such maps of
$[0,1]$ to $P_4(S^1)$ with the same value at $0$ and $1$.
Figure 5 shows
maps like this. The maps shown are continuous, the inverse
images of open sets being open \cite{top}. The map in Fig. 5(a) can be
deformed to the constant map and has zero winding number.The image in
Fig. 5(b) ``winds once around" $P_4(S^1)$. The map in this figure has
winding number one and is not homotopic to the map in Fig. 5(a). It
leads in a conventional way to the generator of ${\bf Z}$. ${\bf Z}$ is
also of course $\p_1(S^1)$.

\sxn{The Laplacian}

We show the construction of the Laplacian $\D$ for the circle poset
$P_{2N}(S^1)$. Its generalization for arbitrary posets will also be indicated.

The numbers $\pm 1$ in Fig. 4(b) are called incidence numbers $I(x_i,x_j)$
\cite{hilton,stanley}. They are attached to each  link $x_i-x_j$
where $x_i$ is regarded as less than $x_j$, $x_i\prec x_j$. We will also
regard $I(x_\a,x_\b)$ as zero if $x_\b$ does not cover $x_\a$.
The boundary operator  $\partial $ can then be defined by
\begin{eqnarray}
\partial x_i & = & \sum _{x_j} I(x_i,x_j)x_j \mbox{ for $i$ odd,}\nonumber \\
\partial x_i & = & 0 \mbox{ for $i$ even, }\label{3.1}
\end{eqnarray}
where the sum in (\ref{3.1}) has a meaning as in homology theory
\cite{hilton}. (\ref{3.1}) reads, in detail,
\be
\partial x_1=x_2-x_{2N},\,\,\,\partial x_3=x_4-x_2,\,\,\,...
\partial x_{2N-1}=x_{2N}-x_{2N-2},\,\,\,\partial x_{2j}=0,\,\,\,
1\leq j\leq N\,. \label{3.2}
\ee
The fundamental property of $\partial $,
\be
\partial ^2 =0 \label{3.3}
\ee
is trivial in this example.

Next let $f_j$ be the characteristic functions supported at $x_j$:
\be
f_j(x_k)\equiv <f_j,x_k>=\d_{jk}\,.\label{3.4}
\ee
[We will use $<f,x>$ and $f(x)$ interchangebly in what follows.]
They are a basis for functions on $P_{2N}(S^1)$ with values in ${\bf C}$. The
boundary operator $d$ on functions is then defined by
\be
<df_j,x_k>=<f_j,\partial x_k> \label{3.5}
\ee
or
\be
df_2=f_1-f_3,\,\,\,df_4=f_3-f_5,\,\,\,...df_{2N}=f_{2N-1}-f_1,
\,\,\,df_{2j-1}=0,\,\,\,1\leq j\leq N\,.\label{3.6}
\ee
It fulfills
\be
d^2=0\,. \label{3.7}
\ee

(\ref{3.6}) shows that $f_j$ can be regarded as forming the ``dual" poset
upside down to Fig.4(b), with $f_{2j}$ at bottom and $f_{2j-1}$ at top, and
with the same incidence numbers on links.

The functions on $P_{2N}(S^1)$ turn into a Hilbert space on introducing the
scalar product
\be
(f_j,f_k)=\d_{jk}\,.\label{3.8}
\ee

Let $d^{\dag }$ be the adjoint of $d$ for (\ref{3.8}):
$$
(d^{\dag }f_j,f_k)=(f_j,df_k)\,\,\,\,\mbox{or}
$$
\be
d^{\dag }f_1=f_2-f_{2N},\,\,\,d^{\dag }f_3=f_4-f_2,...,
d^{\dag }f_{2N-1}=f_{2N}-f_{2N-2},\,\,\,d^{\dag }f_{2j}=0,\,\,\,
1\leq j\leq N .\label{3.9}
\ee
So $d^{\dag }$ behaves like $\partial $. The Laplacian $\D$ is
\be
   \D=-dd^{\dag }-d^{\dag }d\label{3.10}
\ee
or
\begin{eqnarray}
\D f_j & = & f_{j+2} + f_{j-2} -2f_{j},\; \mbox{for $j\neq 1,2,(2N-1)$ and
						$2N $}, \nonumber \\
\D f_1 & = & f_{3} + f_{2N-1} -2f_{1},\nonumber \\
\D f_2 & = & f_{4} + f_{2N} -2f_{2}, \nonumber \\
\D f_{2N-1} & = & f_{1} + f_{2N-3} -2f_{2N-1}, \nonumber \\
\D f_{2N} & = & f_{2} + f_{2N-2} -2f_{2N}. \label{3.11}
\end{eqnarray}
$\D$ looks like the usual discrete Laplacian for each level, but does not
mix the levels.

{}From analogy to homology theory \cite{hilton}, it is plausible that $x_{2j}$
are to be thought of as corresponding to points of the continuum, and
$x_{2j-1}$ as approximations to lines of the continuum
connecting $x_{2j-2}$ and $x_{2j}$.
Thus (\ref{3.6}) can be written as
\be
(df)(x_{2j})=0,\,\,\,\,(df)(x_{2j-1})=f(x_{2j})-f(x_{2j-2});\label{3.12}
\ee
for any function showing that $df$ at $x_{2j-1}$ corresponds to the integral
of $df$ on such a line.
Simplicial decompositions of the posets also suggest a similar idea.
With these interpretations at hand, one can
see that the continuum Laplacian on functions is being approximated here
by the restriction of $\D$ to functions supported at the top, or at the
``rank" zero points, of the poset. [Cf. Section 5.]

The eigenvalues of $\D$ for $P_{2N}(S^1)$ are
\be
\l_k=2(\cos {2k}-1),\; k=m\frac{\p}{N},\; m=1,2,...,N. \label{3.13}
\ee
For the corresponding eigenfunctions $\varphi ^{(k)}$, we have
$$
\D\varphi ^{(k)}=\l_k \varphi ^{(k)},
$$
\be
\varphi ^{(k)}(x_j)\equiv \varphi^{(k)}_j= e^{\imath kj}.
\ee
Since $\varphi _{j+2N}^{(k)}=\varphi ^{(k)}$, they can be regarded as
periodic
functions on the lattice. They are discrete approximations to plane waves
while $\l_k $ are similar approximations to the continuum eigenvalues $-4k^2$.
Note that even and odd $j$ separately furnish eigenfunctions for the same
eigenvalue.

These considerations can be generalised to any poset by introducing
incidence numbers $I(x_\a,x_\b)$ with the following properties \cite{hilton}:
$$
I(x_\a,x_\b)=0\,\,\,\mbox{unless $x_\b$ covers $x_\a$;}
$$
\be
\sum _{x_\b} I(x_\a,x_\b) I(x_\b,x_\g)=0\,. \label{3.15}
\ee
The boundary operator $\partial $ is then defined by
$$
\partial x_\a =\sum _{x_\b}I(x_\a,x_\b)x_\b ,
$$
\be
   \partial x_\a=0 \mbox{ if no point covers $x_\a$, }\label{3.16}
\ee
and fulfills
\be
   \partial ^2=0\,. \label{3.17}
\ee
The operator $d$ on functions is defined using duality as in (\ref{3.5}) and
fulfills the nilpotency (\ref{3.7}) because of (\ref{3.17}). As for the
definition of $d^{\dag }$ and $\D$, we can introduce a scalar product
through
\be
(f_j,f_k)=g_{jk} \label{3.18}
\ee
where $f_j(x_k)=\d_{jk}$. $d^{\dag }$ is the adjoint of $d$ for this scalar
product, while $\D$ is just (\ref{3.10}).

It is desirable to restrict $g_{jk}$ somewhat for physical reasons. For
example, it is reasonable to suppose that $g_{jk}=0$ if $x_j$ and $x_k$
approximate surfaces of different dimension in the continuum or are of
different rank [cf. Section 5].
Locality is also useful for restricting $g_{jk}$. We can
for example set $g_{jk}$ to zero unless $x_j=x_k$ or there is a point $x_l$
linked to both $x_j$ and $x_k$. The simplest choice is of course
$g_{jk}=\d_{jk}$.
\medskip

\centerline{\large \bf 4. The Universal Cover and the Flux}\noindent
\centerline{\large \bf Through  the Circle}
\addtocounter{sxn}{1} \bigskip\medskip

A charged particle on $S^1$ with magnetic flux through its middle can be
described by the free Hamiltonian, but with ``twisted" boundary conditions
(BC's) for energy eigenstates $\psi $ \cite{geons}. They (and their
derivatives)
are not periodic in the coordinate $\varphi $ mod $2\p$ of $S^1$, but change
by $e^{i\theta }$ when $\varphi $ increases by $2\p$.

It is well known \cite{bal} that there is a universal method for the
description of such twists. In this general approach, for the preceding
example, we regard $\psi $ as functions on the universal covering space
${\bf R}^1$ of $S^1$. Here if $x$ is the coordinate on ${\bf R}^1$,
$e^{ix}$ gives the point of $S^1$, so that all points $x+2\p N$
($N\in {\bf Z}$) of ${\bf R}^1$ describe the same point of $S^1$. The twist
is now got by requiring that $\psi (x+2\p)=e^{i\theta }\psi (x)$. This rule
is physically sensible, probability densities being single valued on
$S^1$.

The general method for introducing these twists in continuum physics is as
follows \cite{bal}. If $Q$ is the configuration space, we let $\bar Q$ be its
universal covering space. The fundamental group $\p_1(Q)$ of $Q$ acts freely
on $\bar Q$ and its quotient $\bar Q/\p_1(Q)$ by this action is $Q$. Wave
functions are obtained from
vector valued (one or many component) functions on $\bar Q$
transforming by a unitary irreducible representation (UIR) of $\p_1(Q)$. For
$Q=S^1$, $\bar Q={\bf R}^1$ and $\p_1(Q)={\bf Z}$.

We now show that these ideas work for posets as well. Consider first
$P_{2N}(S^1)$. Its universal cover is just $P({\bf R}^1)$ shown in Fig. 6.
[Its incidence numbers have been obtained from a prescription below.]
$\p_1(P_{2N}(S^1))$ is ${\bf Z}$. It acts according to
$x_i\rightarrow x_{i+2N}$ and the quotient $P({\bf R}^1)$/{\bf Z} is
$P_{2N}(S^1)$.

A systematic approach to find the universal cover $\bar P$ of a poset $P$ uses
the path space $\cp $ as for manifolds \cite{bal}.
$\cp $ consists of continuous maps
of $[0,1]$ to $P$ with the image of $0$ being always a fixed point $x_0$ of
$P$. Let us regard two paths $\G_x$ and $\G_x'\in \cp $ ending at $x$
as equivalent if they are homotopic. A point of $\bar P$ is then the
equivalence
class $[\G_x]$ containing $\G_x$. Finally exactly as for manifolds, $\p_1(P)$
acts by attaching loops to $\G_x$ at $x_0$ and $P=\bar P/\p_1(P)$.

Now $\bar P$ is also a poset. Its poset structure is got as follows: We assume
that $[\G_{x'}]$ covers $[\G_{x}]$ if a) $x'$ covers $x$, and b)
there is a $\tilde \G_{x'}$ in $[\G_{x'}]$ which is a curve
$\tilde \G_x\in [\G_x]$ followed by the point $x'$.
The incidence number for the link from $[\G_x]$ to $[\G_{x'}]$ is taken to be
the same as for the link from $x$ to $x'$.

Let us finally illustrate how to approximate twisted BC's of $S^1$
\cite{geons} using
$P({\bf R}^1)$. The dual $d$ of $\partial $ can be obtained for Fig. 6 in the
usual way. Introducing the scalar product for functions on $P({\bf R}^1)$
through $(f_j,f_k)=\d_{jk}$ where $f_j$ as usual fulfills $f_j(x_k)=\d_{jk}$,
we can also define $d^{\dag }$. The Laplacian $\D$ for $P({\bf R}^1)$ is then
$-dd^{\dag }-d^{\dag }d$. It is easy to check that
\be
\D f_j=f_{j+2} + f_{j-2} -2f_j\,. \label{4.1}
\ee
After imposing the twisted BC's
\be
f_{j+2N}=e^{i\theta }f_j\,,\label{4.2}
\ee
the eigenstates $\chi ^{(k)}$ supported on the top level and the corresponding
eigenvalues $\l_k$ are found to be given by
$$
\chi ^{(k)}(x_j)\equiv \chi _j^{(k)}= A^{(k)}e^{\imath kj}+B^{(k)}
e^{-\imath kj}
$$
\be
\l_k=2(\cos 2k-1) \label{4.3}
\ee
where $k=m\frac {\p }{N}+ \frac{\theta }{2N}$, $m=1,2,3,...,N$.
These are obvious approximations to the continuum answers $-4k^2$ for the
eigenvalues.

\sxn{Solitons}

Topological solitons are characterized by topological invariants such as
winding numbers of maps from a manifold $D$ to a manifold $T$ [8]. The
existence and properties of these solitons are therefore decisively influenced
by the topology of $D$ and $T$. The use of posets, which can approximate
important aspects of topology well, is particularly appropriate for their
study.

The ideal method to approximate the space $F$ of smooth maps from $D$ to $T$
would consist in covering each connected component of
$F$ by a finite number of open sets. We have not
been successful in developing this programme however, $F$ being infinite
dimensional. An alternative to this ideal is to substitute $D$ and $T$ by
posets $P(D)$ and $P(T)$ and consider continuous maps $P^{(0)}(F)$ from
$P(D)$ to $P(T)$, the superscript denoting continuity. We adopt this method
here.

An important feature of this approximation can immediately be pointed out.
Let us
assume that the connected components of $F$ can be labelled by integer
valued winding numbers. If $M$ and $N$ are the number of points of $P(D)$
and $P(T)$, then it is impossible to find winding numbers greater
than $M/N$ in $P^{(0)}(F)$. This is because it is evidently not possible
to cover $T$ by $D$ more often than this ratio.
Thus not all continuum solitons can be realized in this approximation with a
given $M$ and $N$.

We now outline our steps for doing quantum physics using $P^{(0)}(F)$. The
formalism we develop is an approximation to the field theoretic Schr\"{o}dinger
picture, wave functions being functions of fields. There are two parts to
the outline. The first is about the construction of the Laplacian $\D$. $\D$
gives the finite approximation to the continuum kinetic energy. The second
part constructs the finite approximation to the gradient energy [ which
corresponds to an expression like
the spatial integral of $(\partial_i \phi \partial^i \phi)$,
$\phi$ being a field.] Of course if there are other potential energy terms
in continuum dynamics, they too can be finitely approximated: we see no
problem in doing so.

Before going further, we give a convenient characterisation of $P^{(0)}(F)$
which is the set of all
continuous functions from $P(D)$ to $P(T)$. By definition, the
inverse images of open sets are open for continuous functions. It is easily
shown that a function $f$ is continuous iff it preserves order,
that is iff $f(x) \preceq f(y)$ whenever $x\preceq y$.\\

\noindent
{\bf 5.1 The Kinetic Energy}

The first step in finding $\D$ is to convert $P^{(0)}(F)$ into a poset. This
is readily done: we set $f \preceq g$ if $f(x) \preceq f(y)$ for all $x \in
P(D)$.

The next task is to find a $\partial$. Now $\partial$ naturally exists for the
space $P(F)$ of $\underline{all}$ maps from $P(D)$ to $P(T)$ including
discontinuous ones. Let us now explain this $\partial$, its use for our
problem will get clarified later.

A function $f\in P(F)$ is specified by its values $f(i)$ at the points $i$ of
$P(D)$ and can thus be written as $(f(1), f(2), \cdots, f(M))$. [The points
of $P(D)$ can be numbered in any way here.] There being $N$ choices for each
entry, $P(F)$ has $N^M$ points.

The points $y$ of $P(T)$ can be assigned a rank $r(y)$ as follows.
A point of $P(T)$ is regarded as of rank 0 if it converges to no
point, or is a highest point. Let $P_1$ be the poset got from $P(T)$ by
removing all rank zero points and their links. The highest points
of $P_1$ are assigned rank 1. We continue in this way to rank all points.

We will assume that $P(T)$ comes with a boundary operator $\partial _T$
which raises the rank by 1. It is known that such a $\partial_T$ can be
defined if $P(T)$ is a ``rankable" poset \cite{hilton}. What this means is
the following. Starting from a point $y_k$ of rank $k$, we can imagine
travelling upwards in the Hasse diagram along links till we arrive
at a rank zero point. In a rankable poset, the number of  links traversed
on any such path is always $k$. Figure 7 shows the Hasse diagram of a poset
which is not rankable.

The operator $\partial$ for $P(F)$ is defined by
$$
\partial(f(1), f(2),
\cdots, f(M))= (\partial_T f(1), f(2),\cdots, f(M)) + (-1)^{r(f(1))}
(f(1), \partial_T f(2),\cdots, f(M)) + \cdots =
$$
\be
\sum _{j=1}^M (-1)^{\sum_{i<j}r(f(i))}(f(1), f(2), \cdots, \partial _T
f(j), \cdots, f(M)) .
\ee
Clearly, $\partial ^2=0$ as it should be.

There being topological solitons, $P^{(0)}(F) = \bigcup _k P^{(0)k}(F)$
where $P^{(0)k}(F)$ is a connected topological sector. There is no
order relation between elements of $P^{(0)k}(F)$ and $P^{(0)l}(F)$
if $k \neq l$.

Having got $\partial$ for $P(F)$, we may next try to restrict it to
$P^{(0)k}$ for each fixed $k$. But $\partial$ on a continuous
function will in general involve discontinuous functions, namely those
larger than continuous functions. Let $\overline{P}^{(0)k}$ denote the
$\underline {\rm {closure}}$ of $P^{(0)k}$ obtained by adding all these
limit points of $P^{(0)k}$. The $\partial$ operator of (5.1) is then well
defined on $\overline{P}^{(0)k}$.

Figure 8 shows a few elements of $P(F)$ for $D=T=S^1$ and $P(D)=P(T)=P_4(S^1)$.
The arrows on the axes indicate the order relations. The $x_i$ on
abscissa for example can be identified with those in Fig.2 while $y_i$
are points similar to $x_i$ for the image poset. We remark that since
continuous functions preserve order, only those in Figs. 8 (a,b) are
continuous.

These figures show a basic danger in working with $\overline{P}^{(0)k}(F)$. The
discontinuous function there is contained in $\underline {\rm {both}}$
$\overline{P}^{(0)0}$ and $\overline{P}^{(0)1}$ so that different winding
number sectors get mixed up with its inclusion. There is thus the risk of
losing solitons altogether by closing $P^{(0)k}$.

We have shown \cite{sp} that this difficulty does not arise in this example if
$P(T)$ has more than four points. A general result of this kind has also been
developed \cite{sp}.

Having obtained a $\partial $, it is now relatively straightforward to find
$\D$ as explained in Section 3.

We further address such issues elsewhere \cite{sp,int,fedele}.\\

\noindent
{\bf 5.2. The Gradient Energy}

The gradient energy in continuum physics involves differentiation with respect
to the domain variable whereas $\partial$ in (5.1) corresponds to
differentiation in $P(T)$. We must thus invent a new boundary operator
$\hat {\partial}$ to define the gradient energy.

Such a $\hat {\partial}$ is readily available. Let $\partial _D$ be the
boundary operator for $P(D)$ (which we assume exists). If $f \in P^{(0)}(F)$
and $x \in D$, let us write
$f(x)$ as $\langle f,x \rangle$. $\hat {\partial}$ is then defined by
duality:
\be
\langle \hat{\partial} f,x \rangle \equiv
\langle f,\partial_D x \rangle.
\ee
$\partial_D x$ here is an element of the complex vector space generated by
points of $D$, while $f$ is dual to this space with values in the
complex vector space generated by points of $T$.

Next we introduce a scalar product $(^.,^.)$
corresponding to $T$ as in (3.18). $\langle \hat{\partial} f,x \rangle$
being a combination of points of $T$, $(\langle \hat{\partial} f,x \rangle,
\langle \hat{\partial} f,x \rangle)$ is well defined and nonnegative and
can be identified as being proportional to the gradient energy density. The
total gradient energy is obtained by summing over $x$.

All the tools for studying finite approximations to solitons are now at hand.
\\

\sxn{Concluding Remarks}

We have not yet mentioned several important topics for reasons of simplicity.
We will conclude the paper by pointing out a few of these topics and our
own progress in their study.

1) The first topic concerns the utility of the poset approximation for
practical calculations. Definite progress has already been made in its study
\cite{fedele}.
Algorithms for generating soliton posets have been devised and tested.
Numerical work on sine-Gordon solitons are now under way.

2) Another basic subject for research is the poset variant of the Dirac
equation. If $d$ and $\d$ have the conventional meanings of Section 3,
it is known that the familiar Dirac equation in Minkowsky space can be
obtained from $[d+\d+m]\psi=0$ \cite{dirac}. It is also known that a similar
equation on ordinary lattices gives the lattice variant of this equation
\cite{dirac}. As both $d$ and $\d$ make sense on a typical poset, the similar
poset equation seems a good starting point to formulate the poset Dirac
equation.

3) Progress has been made in the study of gauge theories too. When space-
time, and not space alone, is made a poset, we have discovered a
natural way to get the Wilson action [6]. We now outline the idea, leaving
details to another paper. In the absence of gauge fields, the operator $d$
acts on a function $f$ of $P(Q)$ by a rule like $df(i)=f(j)-f(k)$, with $j$
and $k$ covering $i$. Its gauge theory version $\cd$
can be obtained by first introducing the  link variables with
values $U_{ij}$,
$U_{ik}$ in the gauged group $G$, regarded here as unitary matrices acting
on a vector space $V$. With the function $f$ also thought of as valued in $V$,
$\cd $ can be defined using the equation $(\cd f)(i)=U_{ij}f_j-U_{ik}f_k$. It
is
a covariant equation as in lattice gauge theories. Assuming that $U_{ji}=
U_{ij}^{-1}$, we can also show that $(\cd^2f)(i)$ is of the form
$[U_{ij}U_{jk}
U_{kl}U_{li}]f(i)$. The Wilson action is then just a sum over terms of the
form  ${\rm constant} \times Tr[U_{ij}U_{jk}U_{kl}U_{li}]$.

The poset Hamiltonian formulation for $G=U(1)$ also appears to present
no difficulty and can be achieved as follows. $U(1)$ is first
approximated by $P_{2N}(U(1))=P_{2N}(S^1)$. It is also convenient to build
up wave functions from functions on Wilson integrals. For this purpose, we
assign $U_{ij} \in P_{2N}(S^1)$ to each link $ij$ of $P(Q)$, the poset for
the spatial slice $Q$, with $U_{ji}=U_{ij}^{-1}$. [Note that points of
$P_{2N}(S^1)$ can be given a group structure by associating them with
$e^{\imath \frac{\pi }{N}n}$, $n=1,2,...,2N$.]
Now consider continuous
maps of $[0,1]$ to $P(Q)$ with $0$ always having a fixed image $x_0$. By
multiplying $U_{ij}$ along the image of $[0,1]$, we get a map $W$ of $[0,1]$
to $P_{2N}(S^1)$. It is necessary in our approach that $W$ fullfills suitable
continuity conditions which we plan to explain later \cite{sp},
the choice of $U_{ij}$ being accordingly restricted. The space $P^{(0)}(W)$
of these maps is what replaces the space $P^{(0)}(F)$ of Section 5 in  gauge
theories. As it can usually be made a poset just as $P^{(0)}(F)$, we can also
try building Laplacians and other structures appropriate for it.

There is a natural notion of gauge invariance too here. This is because the
group ${\bf Z}_N$ acts continuously on $P_{2N}(S^1)$, the response of
$x_i (\equiv x_{i+2N})$ in Fig.4(b) to its generator being $x_i
\rightarrow x_{i+2}$, and we can think of this group as approximating
$U(1)$. Gauge transformations on $W$ can be defined using it as in lattice
gauge theories\cite{lattice}.

While some of these ideas work for nonabelian $G$, we find difficulties in
conveniently formulating the notion of gauge invariance for such $G$. The
problem is with the symmetries of the approximating posets $P_k(G)$ of $G$.
It is seldom the case that $P_k(G)$ admits a homeomorphism group $G_k$ with
$G_k$ approaching $G$ for large $k$ in some sense. Indeed it is already
impossible to find discrete subgroups $SU(2)_k$ of $SU(2)$ which converge
to $SU(2)$ for large $k$. The treatment of nonabelian gauge theories in the
poset approach thus requires more work.

{\bf Acknowledgements}
\vspace{\medskipamount}

We are very grateful to L. Chandar for numerous discussions and good
suggestions throughout the course of this work. We also thank Ajit Mohan
Srivastava for help with the references.
This work was supported
by the Department of Energy under contract number DE-FG02-ER40231. In
addition, G.B. was supported by INFN(Italy) and P.T.S was
partially supported by Capes (Brazil).

\end{document}